\documentclass{an}
\usepackage{graphicx}
\usepackage{times}
\usepackage{fancyhdr}
\newcommand{\lsim}{\,\lower2truept\hbox{${<\atop\hbox{\raise4truept\hbox{$\sim$}}}$}\,}
\newcommand{\gsim}{\,\lower2truept\hbox{${>\atop\hbox{\raise4truept\hbox{$\sim$}}}$}\,}

\def\tsu27{\left(\frac{T_0}{2.7K}\right)}

\begin{document}

\title{A statistical analysis of a Galactic all sky survey at 1.4 GHz}

\author{C.~Burigana\inst{1}, L.~La~Porta\inst{2}\fnmsep\thanks{Member of 
the International Max Planck Research Shool (IMPRS) of the 
Universities of Bonn and Cologne.}, P. Reich\inst{2} \and W. Reich\inst{2}}

\institute{$^1$INAF-IASF Bologna, Via Gobetti 101, I-40129, Bologna, Italy\\
$^2$Max Planck Institut f\"ur Radioastronomie,
Auf dem H\"ugel, 69, D-53121 Bonn, Germany}

\date{Received; accepted; published online}

\abstract{Radio surveys at 
frequencies of $\sim 1$~GHz allow to map the synchrotron emission (SE) in 
a frequency range where (except for very low Galactic latitudes or 
towards localized regions) it dominates over 
the other radio components.
New all sky total intensity and polarization data at 1.4 GHz have
been recently collected.
We focus on the Galactic radio emission correlation properties 
described in terms of angular power spectrum (APS). We present for the 
first time the APS, in both total intensity 
and polarization modes, for some representative Galactic cuts
and suitable APS power law (PL) parametrizations.
}

\correspondence{burigana@iasfbo.inaf.it}

\maketitle

\vskip -0.15cm
\section{Introduction}
\label{sec:intro}

\vskip -0.15cm
In the recent years a complete coverage of the radio sky at 1420~MHz,
both in total intensity and in polarization intensity, has been 
achieved. It derives from the combination 
of total intensity surveys 
(Reich 1982, Reich \& Reich 1986, Reich et al. 2001)
and of the DRAO 
(Wolleben et al. 2004)
and the Villa Elisa (Testori et al. 2004)
polarization surveys, covering respectively the Northern and Southern 
celestial hemisphere,
sensitive to the Stokes parameters $Q$ and $U$.
All these surveys have a FWHM resolution of $36'$. 
The sky has been sampled at steps of $\simeq 15'$ 
in the total intensity and Villa Elisa surveys.
The preliminary version of the DRAO survey used in this work
has a sky sampling much better than that of 
the Leiden surveys (Brouw \& Spoelstra 1976). 
The typical sensitivities of these surveys 
are of few tens of mK.
These properties allow a reliable study
of the Galactic correlation properties up to a multipole 
$\ell_{max} \sim 250$ ($\ell_{max} \sim 180/\vartheta_{min}(^\circ)$, 
where $\vartheta_{min}$ is the smallest angular scale at which accurate 
information can be extracted).

\vskip -0.3cm
\section{Results}
\label{sec:results}

\begin{figure}
\vskip -0.3cm
\resizebox{\hsize}{!}
{\includegraphics[width=7cm]{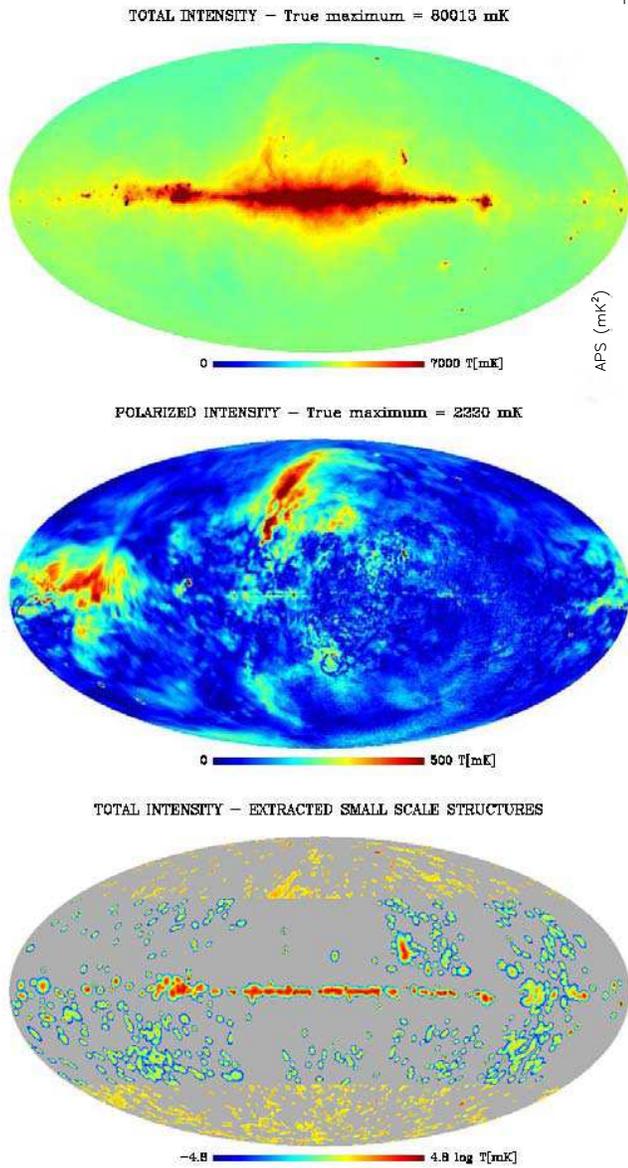}}
\vskip -0.3cm
\caption{All sky maps of the total and polarized intensity at 1420~MHz
(brightness temperature) reprojected according to the 
HEALPix scheme (pixel size $\simeq 6.9'$) and 
of the discrete small scale structures (mainly extragalactic
point-like sources at high Galactic latitudes and Galactic 
point-like or relatively extended sources at low Galactic latitudes)
extracted from the total intensity map (see text).}
\vskip -0.5cm
\label{fig1}
\end{figure}

\vskip -0.2cm
We have properly projected (La Porta et al. 2005) 
these surveys from their original equidistant 
cylindrical pixelization into the HEALPix scheme (G\'orski et al. 2005)
and computed their APS with the 
{\sc Anafast} facility.

Fig.~1 shows the 
maps derived 
from our projection. As evident, the radio sky appears very 
different in total intensity ($T$) and polarization ($PI$):
the polarized intensity on the Galactic plane is relatively 
less bright than in total intensity and exibits important  
structures at high Galactic latitudes. We then
expect a dependence of the APS amplitude 
on the Galactic latitude, $b$, stronger in total intensity than in 
polarization. 

From the HEALPix maps of total and polarized intensity and of $Q$ and $U$ 
parameters we have computed the APS, namely the $T$, $PI$, $E$, and $B$ 
modes, by applying various Galactic cuts. In particular, we exclude in the 
present analysis the region at $|b| < 5^\circ$ where the 
Galactic free-free emission cannot be neglected to focus on the SE
correlation properties.
Fig.~2 shows some representative results. As evident, all the polarization 
modes, and in particular the $E$ and $B$ modes, are very 
similar and 
exibit PL shapes (left panel).
Their typical slopes are in the range $\simeq [-2.5,-3]$, in good 
agreement with the results found by La Porta \& Burigana (2005) in the analysis 
of the Leiden surveys at 1411~MHz. By varying the Galatic cut, the APS 
amplitudes change by a factor $\sim 3$. 
At $\ell \simeq 100$ they are in the range $\sim [0.07,0.2]$~mK$^2$. 
The $T$ mode (right panel) is in general not well approximated
by a single PL but tends to flatten at $\ell \sim 100$
(while the power decreasing at $\ell \gsim 200$ is the obvious 
effect of beam smoothing), particularly at high $|b|$.
To better understand this aspect we performed a dedicated 
subtraction of the discrete structures (DSs) in the total intensity map.
The different level of the diffuse Galactic emission 
at low and high $|b|$ implies a $|b|$ dependent 
source detection level. 
We are confident to have identified DSs (see Fig.~1)
with fluxes above 
$\simeq 1$~Jy (resp. $\simeq 5$~Jy) at $|b| \gsim 45^\circ$ 
(resp. at $|b| \lsim 45^\circ$).
We computed the $T$ APS of the map $i)$ including only the 
detected DSs and 
$ii)$ of the 
diffuse component (plus, obviously, the undetected sources)
derived subtracting DSs
from the original total intensity map.
DSs explain the flattening found at $\ell \sim 100$ 
for Galactic cuts $|b| \ge 5^\circ$ and $|b| \ge 30^\circ$ 
while at high Galactic latitudes ($|b| \ge 70^\circ$) 
they provide the major contribution to the APS also 
at $40 \lsim \ell \lsim 100$.
At $\ell \lsim 100$ (where beam smoothing is negligible) 
the APS of detected DSs is nearly flat 
for $|b| \ge 30^\circ$ and $70^\circ$, as expected in the case of a 
bulk contribution from extragalactic radiosources.
On the contrary, including also the region at 
$5^\circ \le  |b| \le 30^\circ$ the APS of DSs
decreases with $\ell$, 
as expected since the large scale correlation
of the Galactic discrete sources.
Finally, the APSs found for the diffuse component are again 
well described by PLs, with typical slopes
in the range $\simeq [-2.5,-3]$. As expected, the APS amplitude 
strongly depends on the applied Galactic cuts, being not very different
for $|b| \ge 30^\circ$ or $70^\circ$ 
($\approx {\rm some} \times 10^{-1}$~mK$^2$ at $\ell \simeq 100$) but 
significantly larger
($\approx {\rm some}$~mK$^2$
at $\ell \simeq 100$) for $|b| \ge 5^\circ$.

The accurate modelling of the Galactic diffuse SE in both total 
intensity and polarization,
necessary to properly map the intrinsic 
cosmic microwave background anisotropies,
provides relevant information on the Galactic magnetic field 
structure and the cosmic ray electron density distribution.

\vskip -0.15cm
\begin{figure}
\vskip -0.3cm
\resizebox{\hsize}{!}
{\includegraphics[width=5cm]{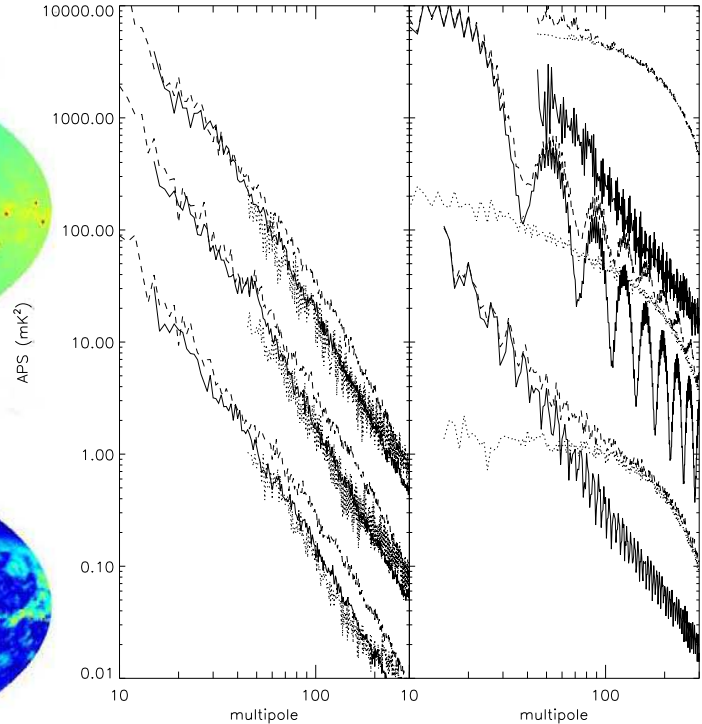}}
\caption{ {\it Left panel}: APS derived from the $PI$ map and from 
the corresponding $Q$ and $U$ maps. The lines starting from $\ell=10$
(dashes), 15 (solid lines), and 45 (dots) refer to $|b| \ge 5^\circ$, 
$30^\circ$, and $70^\circ$, respectively.
From the bottom to the top, each set of three lines refers to
the $PI$ APS, the $E$ mode (multiplied by 10), and
the $B$ mode (multiplied by 100).
{\it Right panel}: APS derived from the $T$ map
for $|b| \ge 5^\circ$ (lines starting from $\ell=10$, APS multiplied 
by 10), $30^\circ$ (lines starting from $\ell=15$), 
and $70^\circ$ (lines starting from $\ell=45$, APS multiplied by $10^3$).
We consider
the original map (dashes), the 
map of the extracted DSs (dots), 
the diffuse component (solid lines).} 
\label{fig2}
\vskip -0.5cm
\end{figure}


\begin{thebibliography}{}
\bibitem{} Brouw, W.N., Spoelstra, T.A.T.: 1976, A\&AS 26, 129
\bibitem{} G\'orski, K.~M., et al.: 2005, ApJ 622, 759
\bibitem{} La Porta, L., Burigana, C.: 2005, A\&A, submitted
\bibitem{} La Porta, L., et al.: 2005, MPIfR, Memo 2005/1
\bibitem{} Reich, W.: 1982, A\&AS 48, 219 
\bibitem{} Reich, P., Reich, W.: 1986, A\&AS 63, 205 
\bibitem{} Reich, P., et al.: 2001, A\&A 376, 861 
\bibitem{} Testori, J.C., et al.: 2004, in: B.~Uyaniker, W.~Reich, 
R.~Wielebinski (eds.), \emph {The Magnetized Interstellar
Medium}, (Katlenberg-Lindau: Copernicus GmbH), p. 57
\bibitem{} Wolleben, M., et al.: 2004, in: B.~Uyaniker, W.~Reich, 
R.~Wielebinski (eds.), \emph {The Magnetized Interstellar
Medium}, (Katlenberg-Lindau: Copernicus GmbH), p. 51
%
\end{thebibliography}
\end{document}